\documentclass[fleqn,usenatbib]{mnras}

\usepackage{newtxtext,newtxmath}
\usepackage[T1]{fontenc}

\DeclareRobustCommand{\VAN}[3]{#2}
\let\VANthebibliography\thebibliography
\def\thebibliography{\DeclareRobustCommand{\VAN}[3]{##3}\VANthebibliography}

\usepackage{graphicx}	% Including figure files
\usepackage{amsmath}	% Advanced maths commands

\usepackage{amssymb}	% Extra maths symbols

\usepackage[dvipsnames]{xcolor}
\title[Search of galactic supernova with ACS/SPI]{Search for gamma-ray emission from a galactic supernova with the anticoincidence system of SPI}

\author[ M. Caixach et al.]{
 M. Caixach,$^{1,2}$\thanks{E-mail: caixach$@$ice.cat}
P. Jean,$^{3,4}$
 J. Isern$^{1,2,5}$
and E. Bravo$^{6}$
\\
$^{1}$Institut de Ci\`{e}ncies de l\textquotesingle Espai (ICE, CSIC), Campus UAB, C/ de Can Magrans s/n, 08193 Cerdanyola del Vall\`{e}s, Spain\\
$^{2}$Institut d\textquotesingle Estudis Espacials de Catalunya (IEEC), C/ Gran Capit\`{a} 2-4, 08034 Barcelona, Spain\\
$^{3}$Universite\'{e} de Toulouse, UPS-OMP, IRAP, 31028 Toulouse, France\\
$^{4}$Institut de Recherche en Astrophysique et Plan\'{e}tologie, 9 Av colonel Roche, BP44346, 31028 Toulouse Cedex 4, France\\
$^{5}$Fabra Observatory, Royal Academy of Sciences and Arts of Barcelona (RACAB), La Rambla 115, 08002 Barcelona, Spain\\
$^{6}$E.T.S.A.V., Univ. Polit\`{e}cnica de Catalunya, c/Pere Serra 1-15, 08173 Sant Cugat del Vall\`{e}s, Spain
}
\date{Accepted XXX. Received YYY; in original form ZZZ}

\pubyear{2021}

\begin{document}
\label{firstpage}
\pagerange{\pageref{firstpage}--\pageref{lastpage}}
\maketitle

\begin{abstract}
The detection of the very early gamma-emission of a Type Ia supernova (SNIa) could provide a deep insight on the explosion mechanism and nature of the progenitor. However this has not been yet possible as a consequence of the expected low luminosity and the distance at which all the events have occurred up to now. A SNIa occurring in our Galaxy could provide a unique opportunity to perform such measurement. The problem is that the optical flux would probably be so attenuated by interstellar extinction that would prevent triggering the observations with gamma-spectrometers at the due time. In this paper we analyse the possibility of using the anticoincidence system (ACS) of the spectrometer SPI on board of the INTEGRAL space observatory for detecting the early gamma-ray emission of a SNIa as a function of the explosion model and distance as well as of pointing direction. Our results suggest that such detection is possible at about 6 - 12 days after the explosion and, at the same time, we can discard missing any hidden explosion during the lifetime of INTEGRAL.
\end{abstract}


%%%%%%%%%%%%%%%%%%%%%%%%%%%%%%%%%%%%%%%%%%%%%%%%%%

%%%%%%%%%%%%%%%%% BODY OF PAPER %%%%%%%%%%%%%%%%%%

\section{Introduction}

Type Ia supernovae (SNIa) are one of the most energetic explosion events in the universe, with luminosity magnitudes comparable to the ones from galaxies. These types of supernova are identified by the lack of H lines and for having strong Si lines in the early spectra \citep{1997Filippenko}. The optical light curve of SNIa peaks at about 20 days after the explosion and decreases 3 magnitudes after one month. They are the consequence of the thermonuclear explosion of a C/O white dwarf (WD) that accretes mass from a companion in a close binary \citep{1960Hoyle}. The discovery of several subtypes of SNIa these last years may be explained by SNIa possibly having more than one kind of progenitor scenarios and explosion mechanisms \citep{2000Hillebrant, maoz2014, 2021Isern}.

Even though the progenitor system and explosion mechanism of such events are not yet proved, one thing is known to be sure about SNIa: their optical luminosities are powered by the decay chain $~^{56}$Ni~$\rightarrow$  $~^{56}$Co $\rightarrow$ $^{56}$Fe.  $~^{56}$Ni is synthesised during the rapid C/O burning and it fuels the explosion by the release of high energy gamma-ray photons that heat the debris and powers the optical and infrared light curves of the event \citep[see][]{1969Colgate}. The detection of gamma-ray lines from this decay can give clues about the kinematics and morphology of the explosion \citep{burr90, gomezgomar1998, J99, 2004Milne, 2008Isern}. The most prominent lines from $^{56}$Ni decay are 158, 480, 750, and 812 keV and from $^{56}$Co decay are 847 and 1238 keV. They can be detected once the ejecta is thin enough in order for the gamma-ray photons to escape without getting thermalized. The decay of $^{56}$Co has a mean lifetime of $\sim$111 days and its lines can be detected right after the maximum of the optical light curve and the following months. The intensity and broadening of the lines give information about the inner layers composition. On the other hand, the decay of $^{56}$Ni has a mean lifetime of $\sim$8.8 days which makes the detection of its gamma-ray lines somehow trickier, as for some explosion models the opacity of the debris is too high for the photons to be able to escape without being absorbed. However, if there are radioactive elements in the outer layers, this gamma-ray emission can give the needed light to the study of the early stage of SNIa. 

The closer SNIa ever detected in gamma-rays is SN2014J. It was discovered by \citet{2014Fossey}. It was found in M82 at a distance of 3.5~Mpc.  SPI instrument on board of INTEGRAL (INTErnational Gamma-RAy Laboratory) collected gamma-rays from day 16.3 to 164 after the explosion. The detection of $^{56}$Ni and $^{56}$Co lines of this nearby event allowed to prove the usefulness of gamma detection to diagnose the dynamics and composition of the ejecta \citep{chur14, dieh14, isern14, chur15, 2016Isern}.
The analysis of these observations  not only confirmed the hypothesis that the light curve was powered by the disintegration of the $^{56}$Ni radioactive chain but also allowed to compute in a direct way the total amount of $^{56}$Ni synthesized during the event. Furthermore they showed, unexpectedly, that $\sim0.05$~M$_\odot$ of $^{56}$Ni was present in the outer layers forming a non-spherical structure. The non-spherical structure was indicated by the display of redshifted and high intensity lines of $^{56}$Ni in the early spectrum.

A Galactic supernova would be ideal for more detailed studies, specially if detected during the rising epoch of the light curve. The expected rate is estimated to be $1.4^{+1.4}_{-0.8}$ events per century \citep{2013Adams} in the case of SNIa, and it is important to realize that the Galactic plane is transparent to gamma-rays for which reason events severely obscured by dust can be detected in gamma. Therefore, taking into account that INTEGRAL has been operating during $\sim 20$ years the probability that such an event has occurred during this period is not negligible.

The anticoincidence system of SPI (ACS) is used not only to reduce the background of SPI but also to monitor the sky for possible sources thanks to its large field of view and its spectroscopic capabilities \citep{2003Vedrenne}. \citet{J99} proposed the use of the SPI-ACS to search for Galactic classical novae, \citet{vonKienlin2001} to locate Gamma Ray Bursts (GRBs), and \cite{RG14} to study solar flares. The scope of this paper is to study the ability of SPI-ACS for detecting Galactic Type Ia supernovae as soon as possible \citep{2019Wang} and, in addition,  to scrutinize the presence of its signature in the already recorded ACS data.

This paper is organized as follows. We present the supernova models used to simulate the detection by the ACS in section~\ref{sec:MODELS}. Section~\ref{sec:SPI} describes the ACS data used for this study. The simulation method and the results of the analyses to estimate the sensitivity of the ACS are presented in section~\ref{sec:SIMULATIONS}. Section~\ref{sec:SEARCH} is dedicated to the search for SNIa signature in the existing SPI/ACS data. Finally, in section~\ref{sec:CONCLUSIONS} we present the discussion and conclusions.

\section{Supernova models}\label{sec:MODELS}
 Two models, DDTe and W7, of SNIa have been used for testing the possibility of detection by ACS/SPI. The spectra as a function of time of these models were obtained with 1D simulations as in \citet{gomezgomar1998}. These models were compared with those obtained from SN 2011fe and SN 2014J by the instruments on board of INTEGRAL \citep{2013Isern,isern14}. 
 Both are compatible with the upper limits deduced for SN~2011fe but the  SN~2014J case demands further considerations since the early spectra presented some features that were better interpreted introducing non-spherical structures. 

\begin{figure}	\includegraphics[width=\columnwidth]{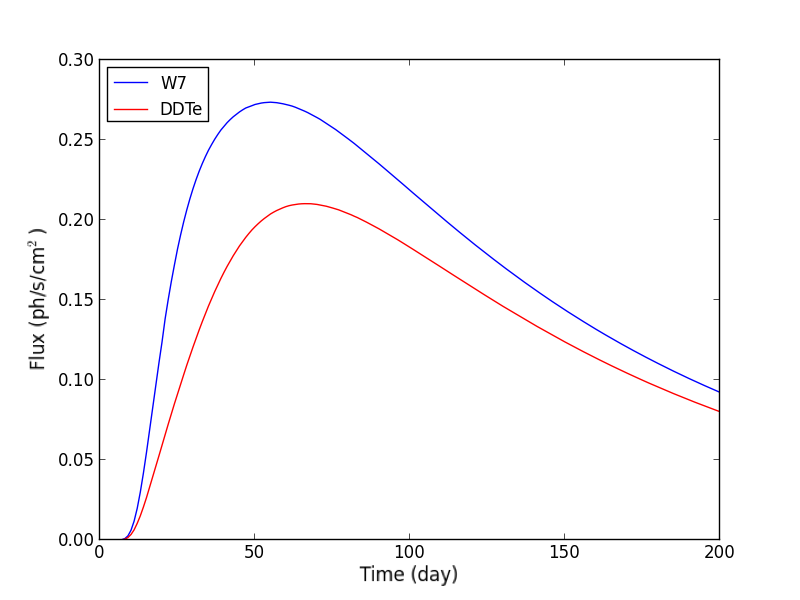}
    \caption{Gamma ray flux variation for model W7 (blue) and DDTe (red) at a distance of 8 kpc from day 5th to 200th after the explosion.}
    \label{fig:LC}
\end{figure}

The DDTe is a sub-luminous SNIa model. It describes a situation in which the burning front starts as a deflagration that turns out into a detonation when reaches a critical density. As a result, material is processed into Fe-group and into intermediate-mass elements. This model ejects 0.51~M$_{\odot}$ of $^{56}$Ni with a kinetic energy of 1.09 $\times$ 10$^{51}$ ergs.
The W7 model is a normal luminous SNIa model \citep{1984Nomoto}. In this model a flame ignited near the center propagates subsonically, thus allowing the expansion of the star and leaving some unburnt material. The W7 model, in particular, is characterized by a small portion of unburnt C-O and a large mass of burnt intermediate-mass elements. This model ejects 0.56~M$_{\odot}$ of $^{56}$Ni with a kinetic energy of 1.24 $\times$ 10$^{51}$ ergs.
Figure \ref{fig:LC} displays the light curve of both models for SNIa at 8 kpc. They have been obtained by integrating the spectrum models from 6 keV to 3684 keV for a SNIa at 8 kpc. 

\begin{figure}	\includegraphics[width=\columnwidth]{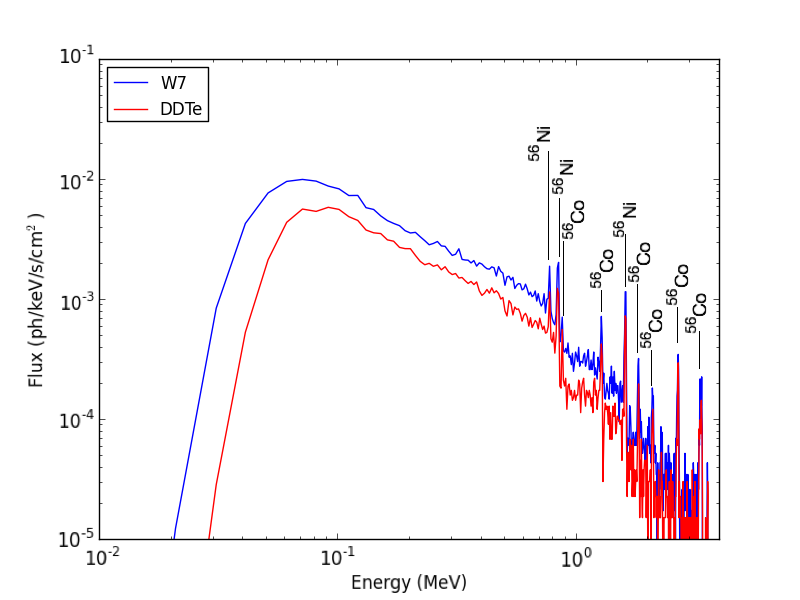}
    \caption{Gamma-ray spectra for day 9th of the models, W7 (blue) and DDTe (red) at 8 kpc. A label on top of each line is added to indicate if they are from $^{56}$Ni or $^{56}$Co decay.}
    \label{fig:spectra}
\end{figure}

Figure \ref{fig:spectra} displays the gamma ray spectra of both models in a logarithmic scale during the 9th day after the explosion.  Even at such early stages the 750 keV and 812 keV  $^{56}$Ni lines decay are visible including several features from the $^{56}$Co decay.   

The slow rising signal of the DDTe model as compared to that of W7 is a convenient property for exploring if the differences in luminosity have an impact on the detection sensitivity.  In order to analyse the detection sensitivity as a function of SNIa distance, models were placed at distances in a range of 4~kpc to 16~kpc. For each model the output flux is provided with an energy resolution of 0.5~keV, in the interval 6~keV to 3684~keV. The spectra were calculated with a frequency of one day from days 5 to 200 after the explosion, although  just the spectra derived for the first days were used and interpolated, as the main goal is to study the detection of the rising signal as soon as possible. For further details about the models data preparation see section \ref{sec:SIMULATIONS}.

\section{The anticoincidence system (ACS) of SPI} \label{sec:SPI}

% Description
The ACS of SPI aims first to reduce the instrumental background in the germanium detectors due to charged particles and gamma-rays coming from outside the spectrometer's field of view. It is composed of an active shield made with 91 scintillator blocks\footnote{Only 89 blocks are active \citep[see][]{Savchenko2012}} in bismuth germanate (BGO) and a plastic scintillator, optically coupled to photomultiplier tubes \citep[see][for a detailed description of the SPI ACS]{vonKienlin2001}. When a particle releases energy in a scintillator block, a veto signal is generated and transmitted to the on-board digital acquisition system. By this way, the active shield allows to remove events produced in coincidence by particles that deposit energy in the shield and in the germanium detectors. The ACS veto rate is recorded with a sampling period of 50 ms and is sensitive to photons releasing energy $\gtrsim$ 80~keV in a scintillator block.

% Use ACS to monitor astrophysical sources
The ACS of SPI is also used to monitor astrophysical sources. For instance, its large detection area allows the detection of gamma-ray bursts and it is part of the INTEGRAL burst alert system providing the  gamma-ray burst location in the sky by triangulations with other space-borne gamma-ray burst monitors {\citep{vonKienlin2001, vonKienlin2003, Savchenko2017}. A giant outburst from the soft gamma-ray repeater SGR 1806-20 was discovered in 2004 by the analysis of the ACS rate \citep{Bor04, M05}.
\citet{RG14} and \citet{G04} explored the capabilities of the SPI ACS to study solar flares.
\cite{J99} proposed to use SPI ACS to search for the hard X-ray and gamma-ray emission from classical novae in our Galaxy \citep[see also][]{Siegert2018}. For this  purpose, they computed the effective area of the SPI ACS with GEANT 3 for energies ranging from 80 keV to 500 keV and a rough angular binning. 

For the present study, we re-calculated the response of the SPI ACS with GEANT 4, using the same INTEGRAL mass model, for energies ranging from 80 keV to 3.5 MeV and with a binning of 10 degrees for the zenithal and azimuthal angles (see Fig.~\ref{fig:effective}). The mass model of INTEGRAL (see Fig~\ref{fig:massmodel}) was developped by \citet{Ferguson2003}. It includes the SPI mass model developed by the SPI collaboration \citep[see][]{Sturner2003}.
For a given photon energy, the response was calculated assuming a point source at infinity (i.e. parallel beam) at a given angular position, by counting the number of events with energy deposit larger than the threshold of 80 keV. The effective area is equal to the number of triggering events divided by the number of simulated photons and multiplied by the exposed area, perpendicular to the parallel beam, in the simulation. 
The values of the effective area obtained in this way are close (within $\sim \pm$30\%) to the previous ones; e.g. $\sim$397 cm$^2$ for E $>$ 100 keV at a zenith angle of 100 deg and an azimuth angle of 0 deg, compared to the value of 340 cm$^2$ close to the SGR 1806-20 position \citep[see][]{M05}. They are also in agreement with the effective areas calculated by \citet{RG14} (e.g. differences are less than 5\%) who used the same INTEGRAL mass model to estimate the gamma-ray flux from some selected solar flares. 

\begin{figure}
	\includegraphics[width=\columnwidth]{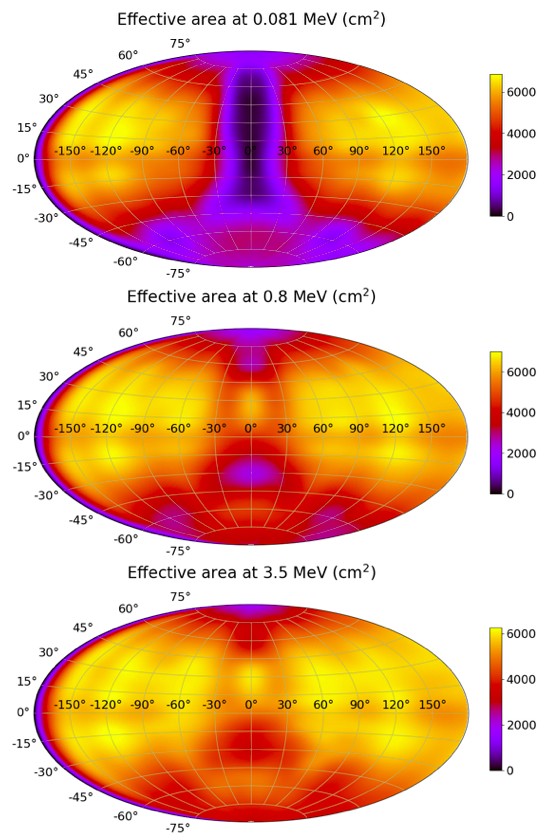}
    \caption{Effective area (cm$^{2}$) of SPI for energies of 0.081, 0.8 and 3.5 MeV for all pointing directions in the sky. The effective area is lower for regions where $z=0$ deg and the energies are low, this is due to SPI being masked by IBIS. }
    \label{fig:effective}
\end{figure}

\begin{figure}	\includegraphics[width=\columnwidth]{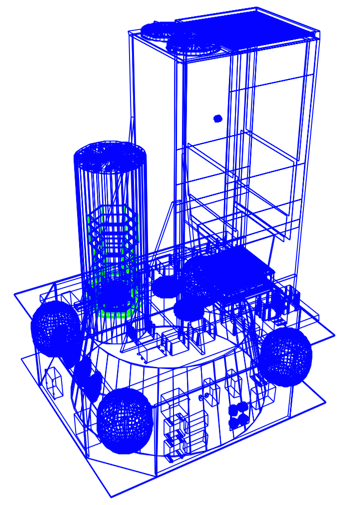}
    \caption{Mass model of INTEGRAL used to calculate the response of the anticoincidence system of SPI (in green) with GEANT 4 (see text).}
    \label{fig:massmodel}
\end{figure}

% Correlation of the ACS saturation event rate with cosmic-ray induced veto rate. 
The anticoincidence system of SPI provides also the saturating event rate, defined as the rate of events released in the BGO blocks with energy  $\gtrsim$ 150~MeV. The ACS saturating event rate, unlike the ACS rate,  is not affected by low energy particles from the radiation belt, from solar flares or from gamma-ray sources that emit mainly in the low energy gamma-ray range  (i.e. E $<$ 100 MeV). Consequently, that makes it a good tracer for monitoring the cosmic-ray intensity at the spacecraft level. When the spacecraft is not affected by these low energy events, the temporal behaviour of the ACS saturating rate is concordant with the total ACS rate. Figure~\ref{fig:rate-vs-year} shows the average ACS rates and the average ACS saturating rates for some revolutions since the launch of INTEGRAL. In order to avoid the effects of radiation belts, the rates shown are the rates averaged over the first to the third quarter of each revolution. Figure~\ref{fig:rate-vs-year} shows that the total and saturating rates follow a linear relation, although the linear coefficients may slightly change due to slight changes of the high energy particle spectrum impinging the ACS.  The saturating rate can be therefore used to trace the variation of the ACS rate without source contribution and be used as background model to our study  (see section \ref{ssec:datafit}).

\begin{figure}	\includegraphics[width=\columnwidth]{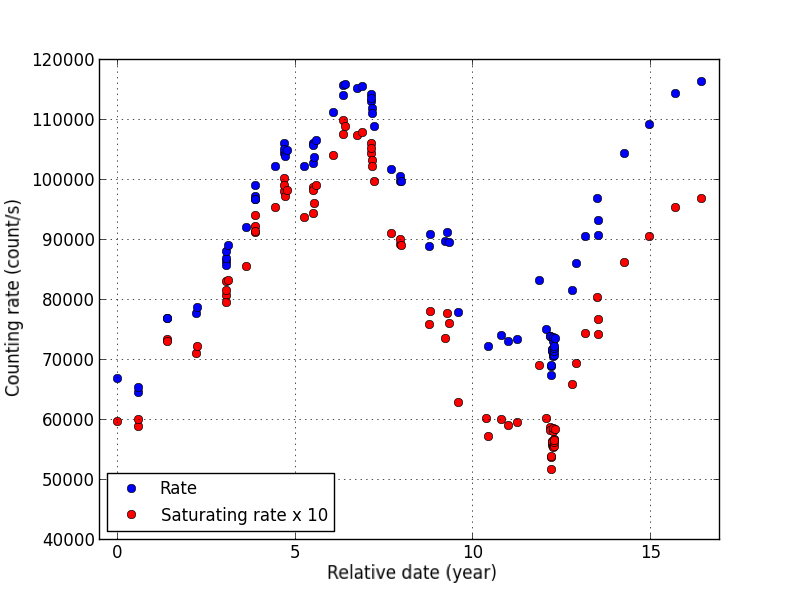}
    \caption{Total and saturating event rates of the anticoincidence system of SPI averaged for some revolutions since the launch of INTEGRAL (see text).}
    \label{fig:rate-vs-year}
\end{figure}

\section{Simulations and analyses}\label{sec:SIMULATIONS}

\subsection{Simulated count rate}
\label{ssec:simulatedcr}

The count rate (CR), counts per second, from a source located at a zenithal and azimuthal angles $\theta$ and $\varphi$ in the SPI frame at the instant $t$ can be defined as:
\begin{equation}
R_{\rm sn}(t) = \sum_{E_{i}} F(E_{i},t) \cdot S_{eff}[E_{i},\theta(t), \varphi(t)] \cdot \Delta E
\label{eq:ratesn}
\end{equation}
where $ F(E_{i},t)$ (counts/cm$^2$/s/keV) is the flux for a given energy and time,  $S_{eff}[E_{i},\theta(t), \varphi(t)]$ is the effective area (cm$^2$ -- i.e. the response matrix) of the ACS and $\Delta$E is the size of the energy bin (keV) of the spectrum. Overall it provides  the rate in counts/s, $R_{\rm sn}$, that would be produced by a SN model. Two SNIa models (see section \ref{sec:MODELS}) have been simulated, the W7 model and a delayed detonation model (DDTe), with an energy binning of  0.5 keV and a time binning of 80 s  to provide enough resolution. Furthermore, the models have been scaled to Galactic distances: 4, 6, 8, 10, 12, 14 and 16 kpc. 

The response matrix has been computed by Monte Carlo simulations (see Section \ref{sec:SPI}). It gives the effective area for energies in the range of 81~keV to 3.5~MeV, 
%(MeV):  0.081, 0.1, 0.12, 0.15, 0.2,  0.25, 0.3, 0.4, 0.5, 0.6, 0.8,  1., 1.5, 2., 2.5, 3., 3.5.
with an azimuthal resolution of 20$^{\circ}$ and a zenithal resolution of 10$^{\circ}$. 

An interpolation of the response matrix is used to take into account the influence of the pointing variation during each revolution on the effective area for each energy event reaching the satellite. INTEGRAL provides for each revolution the right ascension (RA) and declination (Dec) of the x axis of the instrument and the RA and Dec of the z axis. For running the tests we fix the RA and Dec of the simulated source and we take into account the angular evolution of the source with respect to the x and z axis of the instrument to obtain the local coordinates and finally the effective area. 

INTEGRAL provides the ACS data along the lifetime of the satellite for each orbital revolution. The ACS rate from a chosen revolution is added as a background to the SN model to make the simulated count rate realistic. We choose revolutions without strong background variations due to solar flares or early entrance in radiation belts. Each revolution has a duration between 2 to 3 days and the time of each revolution (and the SN model) is binned in intervals of 80 s. Before adding the ACS rate to the SN model rate, the first one needs to be converted from measured rate to true rate by taking into account the dead time $\delta t$ of 0.6  $\mu$s \citep{Savchenko2012}:

\begin{equation}
R_{\rm bck} = \frac{R_{\rm m}}{ R_{\rm m} \cdot \delta t - 1}
\end{equation}
where $R_{\rm bck}$ is the true background rate and $R_{\rm m}$ the measured rate extracted from the data. The rate $R_{\rm sn}$ produced by the SNIa, computed with equation~(\ref{eq:ratesn}), is added to the true background rate to yield to the total rate $R_{\rm tot}$. The final simulated rate $R_{\rm sim}$ is obtained with: 

\begin{equation}
R_{\rm sim} = \frac{R_{\rm tot}}{1 + R_{\rm tot} \cdot \delta t}
	\label{eq:deadtime}
\end{equation}
to take into account the dead time of the ACS.

\subsection{Detection of the supernova signal}
\label{ssec:datafit}

Once the ACS rate produced by the supernova is simulated, we aim to study the sensitivity of the SPI ACS to detect the earliest gamma ray signature of our SN models within a galactic distance.  Two methods have been tested. 

The first one consist of an "ON/OFF". It analyses a previous adjacent window (or interval) in the data, called  the "off" window which represents a null hypothesis, and the subsequent window with the same size, the  "on" window that represents an hypothetical source  \citep{1983Li}. A rising of significance is expected when the "on" window have a rapid increase of flux compared to the previous "off" window. This method, however, has been discarded as the significance of the detections are not high enough. Indeed, when that method is applied, the rate in the "off" period contains the low signal from events produced in the anticoincidence system by the rising gamma-ray emission from the SNIa. The background rate is therefore overestimated. The contribution of the rising signal in the background rate reduces the significance of the detection.

The second method estimates the detection of the rising signal of the supernova by fitting the intensity of the supernova rate and the parameters of a background rate model to the SPI/ACS data. This method takes into account two assumptions. The first one considers that the background rate model (r$_{\rm bck}$) is a linear function of the ACS saturating (R$_{\rm sat}$) event rate :
\begin{equation}
r_{\rm bck} = \frac{R_{\rm sat} - c}{f}
\label{eq:rbck}
\end{equation}
with $c$ and $f$ as free parameters. The saturating event rate (rate ACS saturating) is the rate of events with energy released in the BGO blocks with energies $\gtrsim$ 150~MeV (see section ~\ref{sec:SPI}). The second assumption is that the rate produced by the rising gamma-ray flux of the supernova can be described as the following power law: 
\begin{equation}
r_{\rm sn} = I \cdot (t/5)^{ \beta }
\label{eq:rsn}
\end{equation}
Where $I$ is the intensity and $t$ is the time of the model. Both are free parameters. The power law model is justified by the shape of the gamma-ray light curves when the emission is rising (see Fig.\ref{fig:LC}). The slope of the power law $\beta$ is equal to 18.9 for the W7 model and 14 for the DDTe one. These two different values of the slope does not impact very much on the results.

With these two assumptions we obtain a total model $r_{\rm mod}$ = $r_{\rm bck}$ + $r_{\rm sn}$, where $r_{\rm bck}$ is the background model (see Eq. \ref{eq:rbck}) and $r_{\rm sn}$ is the supernova signal model (see Eq. \ref{eq:rsn}). The best fitting parameters ($I$, $c$ and $f$) are the ones that minimize the chi square calculated as: 

\begin{equation}
\chi ^{2} = \sum \left( \frac{R_{\rm sim} - r_{\rm mod}(I, c, f)}{\sigma_{\rm acs}} \right)^{2} + 10\left( f -f_{0} \right)^{2} + \left( \frac{c}{2000} \right)^{2}
\label{eq:chisq}
\end{equation}
where $R_{\rm sim}$ is the simulated ACS rate (see Eq. \ref{eq:deadtime}) and $\sigma_{\rm acs}$ the standard deviation of the model and the ACS data. The second and third terms in the sum are penalty functions that avoid wrong best fitting values of the parameters $f$ and $c$. Without those penalty functions, the minimization process could converge toward local chi square minimum yielding to abnormal values of the parameters of the background model, $f$ and $c$. It sometimes occurs when the signal of the supernova is strong. The parameter $f_{0}$ in Eq. \ref{eq:chisq} is an estimated mean value of the factor that scales the ACS total rate with the ACS saturated rate (see Fig. \ref{fig:rate-vs-year} and section \ref{sec:SPI}). The factors 10 and 2000 in Eq. \ref{eq:chisq} were chosen to limit the variations of the f and c on a reasonable range of values.

From these results we can estimate the detection date. To do so, we estimate first the quality of the background model ($r_{\rm bck}$) computing a residual ($\delta$): 
\begin{equation}
\delta=\frac{R_{\rm sim}-r_{\rm bck}}{\sigma_{\rm acs}}
\end{equation}
Then, this  residual is smoothed to avoid a false detection due to a single short time scale spike in the data.
 
We consider that the supernova signal is detected when the following conditions are satisfied: 1) t$_{\rm acs}$ $>$ t$_{\rm start}$ + 0.5~day;  2) $\delta >$ 3.
Where  t$_{\rm acs}$ is the time of the data, the sigma level is set to 3, and  t$_{\rm start}$ is the date at which the ACS data starts to be recorded at the beginning of an orbital revolution of INTEGRAL}. The addition of half a day in the first condition aims to remove significant count rate excess induced by high energy particles when INTEGRAL goes out of the radiation belts, after its perigee.

This procedure is done to test the DDTe SN model and the W7 SN model over 275 revolutions. These revolutions have been taken along the lifetime of INTEGRAL to find the earliest time of detection. Each of them have the following available data: ACS rate, ACS saturating rate, rate time, right ascension and declination of the INTEGRAL X-axis, right ascension and declination of the INTEGRAL Z-axis. The ACS rate is used as simulated background to the models, as mentioned in section  \ref{ssec:simulatedcr}. The ACS saturating event rate is used on this analysis to estimate the background ACS rate and detect the supernova signal, as mentioned previously in this section.  Each revolution has a length of 2 to 3 days that we fixed to $\approx$2 days as we cut the initial hours of data because most of them still contain some noise from the radiation belt phase\footnote{SPI is switched off between revolutions}.  A same length of the supernova model is cut in order to compute the simulated count rate. The initial time of the computed count rate equals to a time ranging from day 4th to 6th of the SNIa model, for earlier days the flux's model is not high enough to make a contribution to the ACS rate. This initial time, from 4th to 6th, is chosen randomly to give more realistic circumstances to the analysis.

\begin{figure}	\includegraphics[width=\columnwidth]{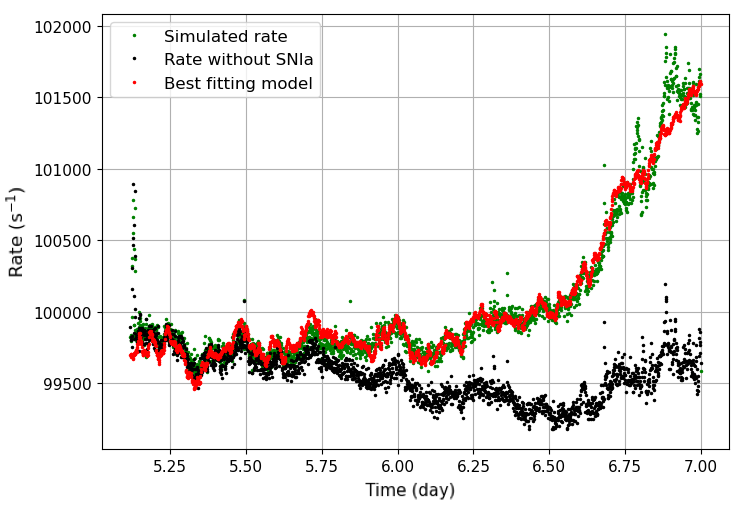}
    \caption{Display of the SN model rate with revolution number 1752 added as background in green. The background rate, without the SN signal, is shown in black. The best fitting model to the SN with background rate is shown in red. For this example, the rising of the SN rate is appreciated and allows the detection time to be at 6.73 day.  }
    \label{fig:acsfit}
\end{figure}

When the count rate with a length of 2 days is analyzed two things may happen: a supernova signal detection is found or not. If a detection is found in the count rate then the revolution number, the time detection and its significance are saved. In case it is not, we proceed to analyze the consecutive 2 days of the computed count rate, which would account for the 2 following days of the SNIa model and using the subsequent revolution as its background. This procedure is repeated until a detection is found or until we reach an initial time of the count rate higher than the 10th of the SNIa model. When this point is reached, we start again the process with an initial count rate time ranging from 4th to 6th but with a different revolution as background and so on. An example output of this method is shown in Figure ~\ref{fig:acsfit}, where the count rate is computed using the DDTe model with revolution number 1752. The computed discovery date is at 6.73 day with  $\approx$ 67$\sigma$, which is coherent if we make a visual evaluation of the count rate data.  

In some cases, due to high background or too many changes on pointing direction in the same revolution, the rising of the SN model goes undetected but on the following revolution, as the SN model flux is higher with time, the flux is too high to find a good fit. In these cases an inspection of the data makes clear that there is a high rising signal and the satellite would recognize it as such. We want to count these cases as late detections, therefore a condition is created where, if the maximum of the SNIa signal is larger than 8.5$\%$ of the mean ACS rate, the detection time is saved even if the fit is not good enough. 

The analysis of the count rate has been performed for each of the models (DDTe and W7) for distances of 4, 6, 8, 10, 12, 14 and 16 kpc. For each distance and model several pointing directions have been tested to reproduce different galactic areas backgrounds and to check the variability of the detection  on different positions in the Galaxy. The pointing directions, in galactic coordinates of longitude (l) and latitude (b), are l = 0, 45, 90 and 180 deg for b = 0 deg. All the position have a same value of latitude of b = 0 deg. The reason for this fixed latitude is that the location for these events in our galaxy would be in regions of old population stars, which are in the old disk and in a spheroid centered to the galactic center. Therefore, they would likely occur at $\pm$ 15 deg from the galactic plane. As seen in Figure \ref{fig:effective}, the response of the ACS does not change significantly with the direction of the source (except for low energies). Hence, simplifying the latitude to b = 0 deg is good enough for our study.

\subsection{Results}
\label{ssec:results}

\begin{figure}
	\includegraphics[width=\columnwidth]{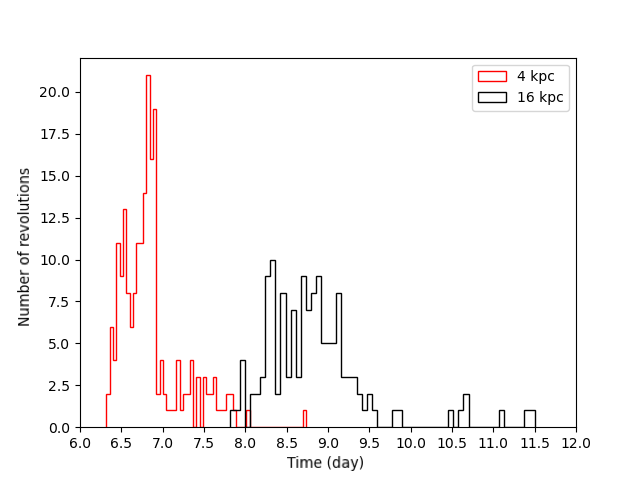}
	\includegraphics[width=\columnwidth]{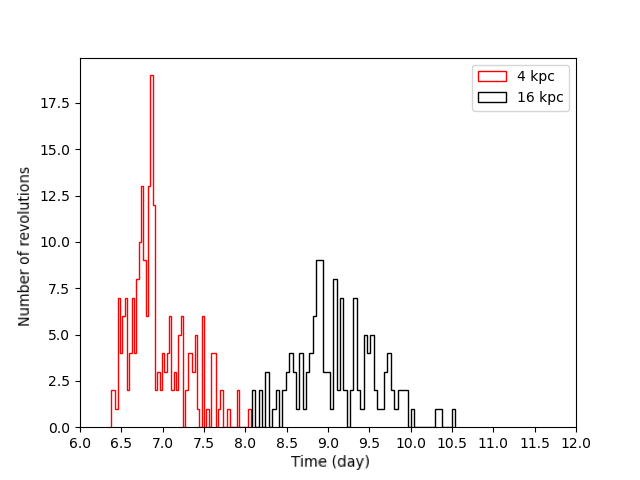}    
    \caption{Top panel shows the distribution of the discovery dates for the W7 model at l = 0 deg and b = 0 deg for distances of 4 kpc (red) and 16 kpc (black). Bottom panel displays the same for the DDTe model.}
    \label{fig:hist}
\end{figure}

Following the methodology explained on Section~\ref{ssec:datafit}, the time detection for the W7 and DDTe models placed at  distances of 4, 6, 8, 10, 12, 14 and 16 kpc and at positions l = 0, 45, 90 and 180 deg with b = 0~deg has been computed. An example of two different distances, 4 kpc and 16 kpc, and l, b = 0 deg, is shown for models W7 and DDTe in Figure~\ref{fig:hist}, top and bottom respectively. For both models, the number of detections for higher distances is lower due to the lower flux. 

\begin{figure}
	\includegraphics[width=\columnwidth]{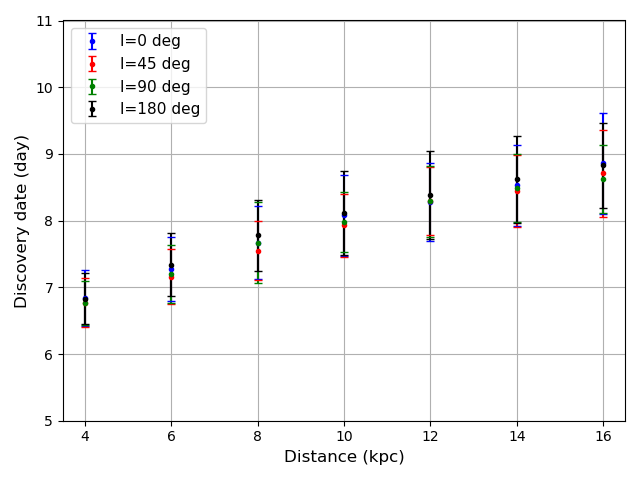}
	\includegraphics[width=\columnwidth]{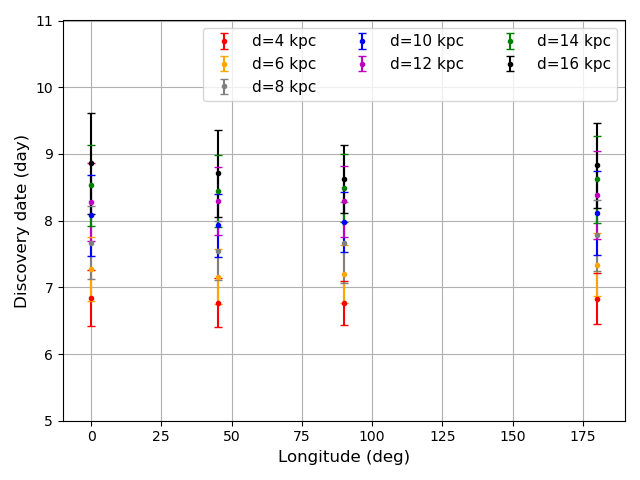}    
    \caption{Mean discovery dates versus distance for model W7 (top panel) and  longitude (bottom panel).}  
    \label{fig:longw7}
\end{figure}

\begin{figure}
	\includegraphics[width=\columnwidth]{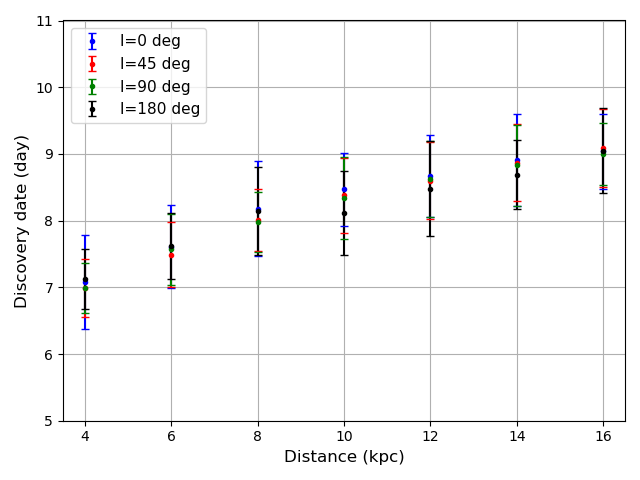}
	\includegraphics[width=\columnwidth]{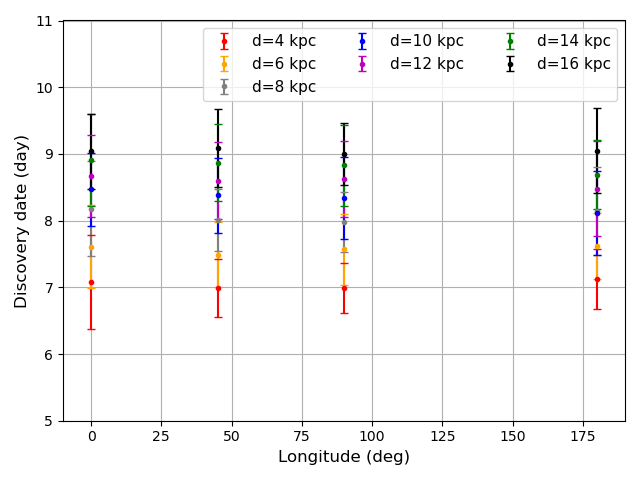}   
    \caption{Mean discovery dates versus distance for model DDTe (top panel) and  longitude (bottom panel). } 
    \label{fig:longddt}
\end{figure}

The mean value of the distribution is computed and shown for
all distances and pointing directions on Figure~\ref{fig:longw7} for the W7 model and Figure~\ref{fig:longddt} for the DDT model. The time of the early signal detection for each distance is coherent with the flux $\propto$ distance$^{-2}$ relation, as for further distances the detections are later due to lower flux. The W7 model has a slightly earlier detection for each case because of its faster rise compared to the DDTe one. The position of the source is not relevant for the early detection of the model, as it does not change for each longitude tested, see bottom Figure~\ref{fig:longw7} and~\ref{fig:longddt}. These results are due to the fact that INTEGRAL observatory has been pointing in various directions during the $\sim$15 years of observation and that the equivalent field of view of the ACS/SPI is very large, therefore the position of a SNIa in the sky does not show any effect. 

%\section{Search of long gamma-ray signature in the SPI/ACS data}\label{sec:SEARCH}
\section{Search of SNIa signature in the SPI/ACS data}\label{sec:SEARCH}

We performed a systematic search for a SNIa signature in the ACS rate with the data available at the date of the analysis: 1868 revolutions from revolution 26 to revolution 2063. The analyses were made for each revolution using ACS data rebinned with a sampling period of 18~s. The detection of a SNIa signature is triggered when the smoothed residuals\footnote{The residuals were smoothed with a hanning window of 1.5~h to reduce the number of detection due to gamma-ray bursts and false detection due to statistical fluctuations.} obtained after subtraction of the background model (see section 4.2), are larger than 5$\sigma$ up to the end of the analyzed revolution. This criterion, which was chosen for its simplicity, allows the detection of a flux rising with a long timescale as to the one expected from a SNIa (e.g. see Fig. \ref{fig:acsfit}). A first analysis of all data yielded to a lot of detection due to background variations, which spans up to the end of a revolution, that were not well modeled with the ACS saturating event rate. Those background variations are due to low energy particles of the radiation belt or from solar events. The signature of such particles are clearly seen in the data of the INTEGRAL Radiation Environment Monitor (IREM), which measures the flux of particles with energy $>$ 0.5~MeV for electron and 10~MeV for proton \citep{Haj03}. The counter TC3 of the IREM reproduces well the variations of ACS rate produced by low energy charged particles of radiation belts. In order to reduce the number of detections due to low energy particles, we added the rate of the TC3 counter in the background model for this specific analysis.  

\begin{figure}
\begin{center}
\includegraphics[scale=0.5,angle=0]{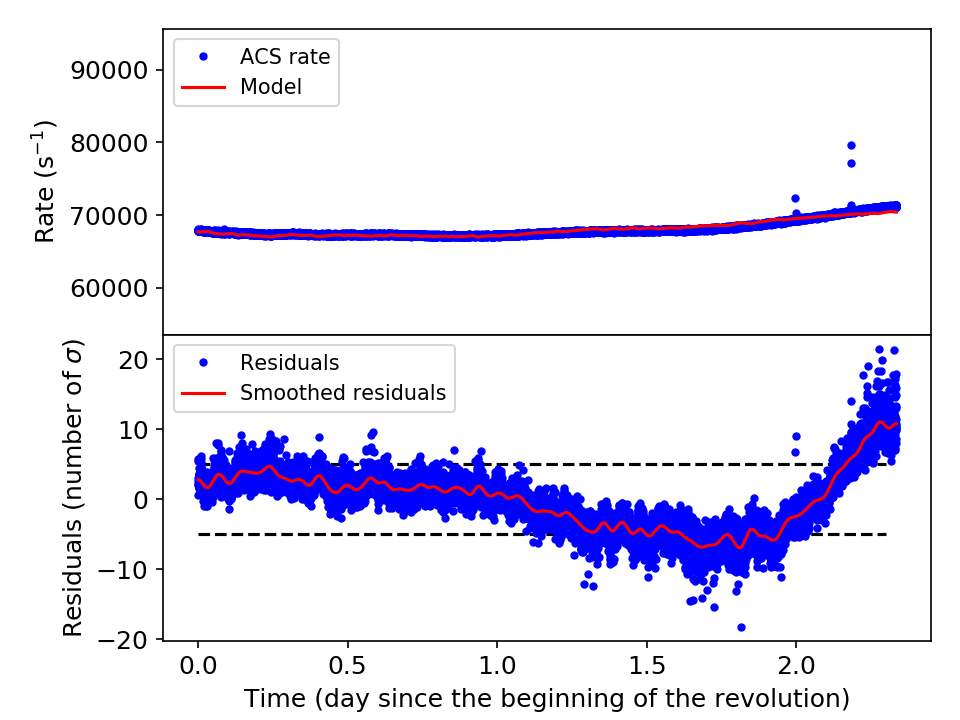}
\caption{Example of measured and modeled ACS rates (top) and the residuals (bottom) for the revolution 1342. The imperfect modeling of the ACS rate at the end of the revolution, when INTEGRAL enters into radiation belts, produces a smoothed residual excess larger than 5$\sigma$.}
\label{fig:ex_rev1342}
\end{center}
\end{figure}

With such a detection method, we counted 58 revolutions with SNIa signature. However, all of them are detected at dates larger than $\sim$ 2.2~days after the beginning of the revolution. These excesses are explained by an imperfect modelling of the enhancement of the rate produced by low energy particles impinging on the scintillators of the ACS when INTEGRAL enters into the radiation belt at the end of its orbital revolutions (e.g. see Figure \ref{fig:ex_rev1342}). 

% Figure 

\section{Conclusions}\label{sec:CONCLUSIONS}
This study demonstrates the capability of the ACS/SPI to detect a galactic supernova during its early stages. Two theoretical SNIa models with different luminosities rising times, W7 and DDTe, have been used in order to evaluate the  influence on the time of the detection. For such purpose we have developed a method to detect the signature of the rising flux emitted by a supernova and to derive a detection date as quickly as possible. 
%By using the count rate data of the modelled sources and fitting them to the background ACS rate of 275 revolutions, 
Using simulations of the count rate produced by the source models and the measured background ACS rates of 275 orbital revolutions of INTEGRAL, we have found that our method allows to discover a galactic SN as soon as 6 days up to 12 days after explosion, depending of the distance and model. 
% please note that the latest detection date in Fig. 6 is 11.5 days. So formally, 9 days should be replaced by 12 days.

The simulated sources have been placed at several distances (from 4 to 16 kpc), several galactic longitudes and zero latitude. The discovery date increases with the distance, as the flux of the modelled sources is smaller. This result follows the distance flux relationship. On the other hand, the galactic longitude of the source has no influence on the discovery date, which is logical taking into account the large field of view of the ACS/SPI.
%We argue that these results are due to the large field of view of the ACS/SPI. 

As the discovery date depends on the flux, we have found there is a slightly different range of discovery dates for the two models. The DDTe model is a sub-luminous SN Ia model and the W7 a normal luminous one. The lower luminosity of the DDTe resulted in a detection range of approximately half-day later than the W7.
Using our detection method we have analyzed all the available ACS/SPI data from the beginning of the INTEGRAL mission to now  and we have not found any signature of a galactic SNIa, suggesting no event has been missed.

This method is not able to provide a position of a source but it can provide an early alert to activate other in orbit or ground-based observatories to quickly browse the sky in search for it. In any case, the collected data would provide very important information about the early development of the supernova outburst no matter if the event is detectable in any other energy range. INTEGRAL is on its last stages of life and the results obtained here suggest that future missions should consider the possibility of using the ACS as an all sky detector by improving as much as possible its sensitivity. The new space telescope COSI \footnote{see website: https://www.nasa.gov/press-release/nasa-selects-gamma-ray-telescope-to-chart-milky-way-evolution}, which is expected to launch in 2026, will observe the gamma-ray sky in the 200 keV - 5 MeV range \citep[]{Tomsick2021}. It will be equipped of a compact Compton telescope made with germanium detectors and of an anticoincidence system made with BGO scintillators as for SPI. With its large field of view ($\sim$ 25 \% of the sky) and its pointing strategy, COSI will monitor the entire sky within a day. Both the Compton telescope and the anticoincidence system will be sensitive to the gamma-ray emission of a galactic SNIa.

\section*{Acknowledgements}
The authors thank the anonymous referee for helpful comments. We acknowledge support from the Spanish Ministry of Science and Innovation and - FEDER UE (MCI-AEI-FEDER,UE) through grants PID2019-108709GB-I00 (MC,JI) and PGC2018-095317-B-C21 (EB), by grant 2014 SGR 1458 and CERCA Programe of the Generalitat de Catalunya (MC,JI), and  by the program Unidad de Excelencia Mar\'{i}a de Maeztu CEX2020-001058-M (MC,JI).

\section*{Data Availability}
The SPI data used for this study are public data available in the ISDC website: \url{http://www.isdc.unige.ch/integral/archive#DataRelease}.


\bibliographystyle{mnras}
\bibliography{references} % if your bibtex file is called example.bib

%\begin{thebibliography}{99}
%\end{thebibliography}

%%%%%%%%%%%%%%%%%%%%%%%%%%%%%%%%%%%%%%%%%%%%%%%%%%

%%%%%%%%%%%%%%%%% APPENDICES %%%%%%%%%%%%%%%%%%%%%

%%%%%%%%%%%%%%%%%%%%%%%%%%%%%%%%%%%%%%%%%%%%%%%%%%

% Don't change these lines
\bsp	% typesetting comment
\label{lastpage}
\end{document}